\documentclass[sigplan,nonacm]{acmart}

\received{January 2023}
\graphicspath{ {./fig/} }
\usepackage{units} % nice fractions \nicefrac
\usepackage{subcaption} % sub-figures
\usepackage{varioref} % nice pronounced references, use \vref \vpageref

\newcommand{\real}{\mathbb{R}} % real numbers
\newcommand{\degree}{\text{\textdegree}} % degrees in math
\allowdisplaybreaks[1]

\theoremstyle{remark}

	\newtheorem*{problem*}{Problem}
	\newtheorem*{prop*}{Proposition}
	
	\newtheorem*{example*}{Example}
\theoremstyle{definition}
	
\theoremstyle{plain}

\begin{document}

\title{Radiance Textures for Rasterizing Ray-Traced Data}

\author{Jakub Maksymilian Fober}
\orcid{0000-0003-0414-4223}
\email{talk@maxfober.space}

\renewcommand\shortauthors{Fober, J.M.}

\begin{abstract}
% Intention
	Presenting real-time rendering of 3D surfaces using radiance textures for fast synthesis of complex incidence-variable effects and environment interactions. This includes iridescence, parallax occlusion and interior mapping, (specular, regular, diffuse, total-internal) reflections with many bounces, refraction, subsurface scattering, transparency, and possibly more.
% Method
	This method divides textures into a matrix of radiance buckets, where each bucket represent some data at various incidence angles. Data can show final pixel color, or deferred rendering ambient occlusion, reflections, shadow map, etc. Resolution of the final synthesized output is the radiance bucket matrix size. Technique can be implemented with a simple fragment shader.
% Oucome
	The computational footprint of this technique is of simple diffuse-only graphics, but with visual fidelity of complex (off-line) ray-traced render at the cost of storage memory footprint.
% Discussion
	Balance between computational footprint and storage memory footprint can be easily achieved with variable compression ratio of repetitive radiance scene textures.
\end{abstract}

%%%%%%%%%%%%%%%%%%%%%%%%%%%%%%%
%% CSS generated code        %%
%% http://dl.acm.org/ccs.cfm %%
%%%%%%%%%%%%%%%%%%%%%%%%%%%%%%%
%%%%%%%%%%%%%%%%%%%%%%%%%%%%%%%%%%%%%%%%%%%%%%%%%%%%%%%%%%
%% The code below should be generated by the tool at    %%
%% http://dl.acm.org/ccs.cfm                            %%
%% Please copy and paste the code instead of one below. %%
%%%%%%%%%%%%%%%%%%%%%%%%%%%%%%%%%%%%%%%%%%%%%%%%%%%%%%%%%%



%%%%%%%%%%%%%%%%%%%%%%%%%%%
%% End of generated code %%
%%%%%%%%%%%%%%%%%%%%%%%%%%%


\maketitle

%%%%%%%%%%%%%%%%%%%
%% Small license %%
%%%%%%%%%%%%%%%%%%%

\begin{figure}[bh]
	% Creative Commons 3.0 logo
	\footnotesize
	\copyright\ \the\year\ Jakub Maksymilian Fober\smallskip\\
	\hyperlink{https://creativecommons.org/licenses/by-nc-nd/3.0/}{\includegraphics{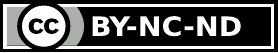}}\\
	This work is licensed under Creative Commons BY-NC-ND 3.0 license.
	\url{https://creativecommons.org/licenses/by-nc-nd/3.0/}\\
	For all other uses including commercial, contact the owner/author(s).
\end{figure}

%%%%%%%%%%%%%%%%%%
%% Introduction %%
%%%%%%%%%%%%%%%%%%

\section{Introduction}

% Field of study
Radiance and reflectance field rendering techniques are a class of algorithms used in computer graphics to generate images of three-dimensional scenes. These algorithms simulate the way light interacts with surfaces in a virtual environment, producing realistic and detailed images.

These techniques have been the subject of extensive research in computer graphics and rendering, as they offer a powerful and flexible way to generate high-quality images. There is a wide range of applications for radiance and reflectance field algorithms, including film and video game production, architectural visualization, and scientific visualization.

% Hypothesis
In this paper, technique is presented to capture and render complex precomputed light interactions, via radiance field textures, embedded onto three-dimensional-object's surface.

% Applicability
The presented technique utilizes a standard fragment pixel shader and a two-dimensional texture lookup to render dynamic, view-independent, photo-realistic images at a fraction of the computational cost associated with effects such as real-time ray tracing, parallax mapping, and dynamic shadowing.

It is well-suited for real-time execution in video games, virtual reality, and virtual production environments on modern hardware. It can take advantage of the direct storage capability in ninth-generation gaming systems, providing high-fidelity, high-performance images.

% Do not change
This technique can replace computationally heavy rendering-pipeline chains, while preserving hardware-accelerated, highly-optimized rasterization elements. It can also enable wider implementation of real-time GPU ray-tracing, with ability to combine bounce rays with precomputed radiance of the environment.

% Compare to existing methods
\subsection{Previous work}

Mainstream implementations of radiance field rendering focus on volumetric data structures and spherical harmonics for rendering images\cite{Yu2021Plenoxels}. While volumetric data can be sparse in order to exclude void regions\cite{Yu2021Plenoxels}, the ultimate goal would logically be to perfectly match the geometry of the represented object. And since the inside volume of the object is of no interest (most of the time), only half of the radiance sphere is considered practically useful. Therefore, such fields could effectively be spread across the surface of the object.

Some researchers embraced this approach, with neural reflectance fields as texturing primitives\cite{Baatz2022NerfTex}, which rendered high-fidelity results. But while neural fields produce fantastic results, they are computationally intensive at rendering time\cite{Yu2021Plenoxels} and therefore are not suitable for real-time applications.

% Overview of th paper
\subsection[Overview]{Overview of the content}

In this initial version of paper you will find theoretical explanation and implementation of the subject, along with equations and schematics. Some elements had been tested, like mapping functions, some yet to be presented, as the follow-up updates continue.

% Introduction to document (technical)
\subsection[Naming convention]{Document naming convention}

This document uses the following naming convention:
\begin{itemize}
	\item Left-handed coordinate system.
	\item Vectors presented natively in column.
	\item Row-major order matrix arranged, denoted ``$M_{\text{row}\,\text{col}}$''.
	\item Matrix multiplication by ``$[\text{column}]_a\cdot[\text{row}]_b=M_{a\,b}$''.
	% \item Double bar enclosure ``$\Vert\vec A\Vert$'' represent vector direction.
	\item A single bar enclosure ``$|u|$'' represents scalar absolute.
	\item A single bar enclosure ``$|\vec v|$'' represents vector's length.
	\item Vectors with an arithmetic sign, or without, are calculated component-wise and form another vector.
	\item Centered dot ``$\cdot$'' represents the vector dot product.
	% \item Cross sign ``$\cross$'' represents the vector cross-product.
	\item Square brackets with a comma ``$[f,c]$'' denote interval.
	\item Square brackets with blanks ``$[x\ y]$'' denote vectors and matrices.
	\item The power of ``$^{-1}$'' implies the reciprocal of the value.
	\item QED symbol ``$\square$'' marks the final result or output.
\end{itemize}
This naming convention simplifies the process of transforming formulas into shader code.

% How it works
\section{Methodology}

Each pixel of the model's texture contains discrete radiance hemispherical map of size $n\times n$, called ``bucket''. Buckets are arranged in place of initial texture's pixels, increasing overall resolution to $w\cdot n\times h\cdot n$ pixels, where $w$ and $h$ denote \emph{width} and \emph{height} of the synthesized output texture, respectively. Buckets are highly repetitive and change only slightly from one to another. This is a great case for a simple compression.

To synthesize output texture for a given view position, single sample per bucket is taken, giving normal resolution texture output.

Model's $u,v$ texture coordinates correspond to bucket matrix position index, while incidence vector, correspond to bucket's internal $u,v$ position. Therefore radiance texture sampling algorithm can be described as a four-dimensional plenoptic function $L(u,v,\theta,\phi)$, where $u,v$ denote model's texture coordinates and $\theta,\phi$ incidence angles.

\begin{figure}[h]
	\includegraphics{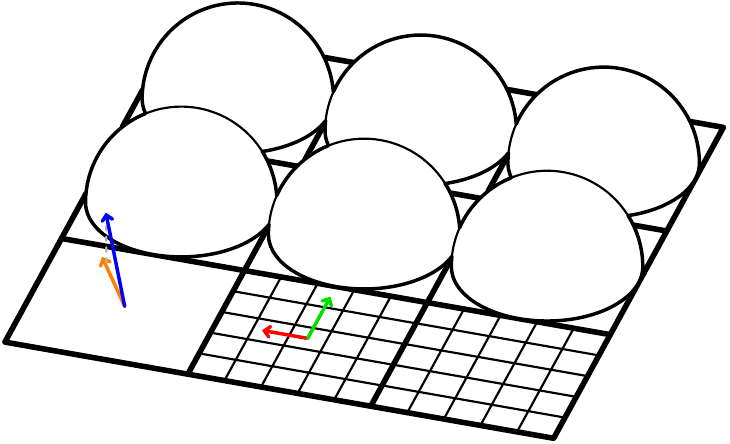}
	\caption{Radiance texture sampling model, where the incidence $\real^3$ vector \emph{(blue)} is projected and squarified \emph{(orange)} to $\real^2$ texture coordinates \emph{(red} and \emph{green)}, which map onto hemispherical radiance bucket represented as a flat square.}
	\label{fig:main drawing}
\end{figure}

Each radiance bucket should represent a hemisphere of reflectivity. Equisolid azimuthal projection was chosen for this task, for its properties, as it preserves area and resembles spherical mirror reflection\cite{wikiFisheyeLens}.
Resolution of the radiance bucket, in such projection, directly corresponds to $\sin(\nicefrac{\theta}{2})\sqrt{2}$, where $\theta$ is the incidence angle.

To efficiently spread information across square buckets, additional disc-to-square mapping function was implemented, providing uniform pixel count across both orthogonal directions and diagonal directions.

Equisolid azimuthal projection mapping can be easily implemented in the vector domain without the use of anti-trigonometric functions, as the orthographically projected normalized sum of the incidence and normal vectors has a length of $\sin(\nicefrac{\theta}{2})$. This eliminates $\theta,\phi$ angles from the plenoptic function, resulting in new $L'(u,v,x,y,z)$, where $x,y,z$ correspond to incidence unit-vector components in orthogonal texture space.

\subsection[Incidence to texture]{Mapping of incident vector to radiance bucket}

For every visible pixel there is an incidence vector $\hat I\in\real^3$.
This vector can be mapped and projected to $\real^2$ texture coordinates using translation and $\real^{2\times3}$-matrix transformation.

Following equation maps incidence vector to azimuthal equisolid projection, with $r=1$, at $\Omega=180\degree$.
\begin{subequations}
\begin{align}
	\begin{bmatrix}
		\vec A_x \\
		\vec A_y \\
		\sqrt{2}\cos\nicefrac{\theta}{2}
	\end{bmatrix}
	&= \sqrt{2}
	\left\|
	\begin{bmatrix}
		\hat I_x + \hat N_x \\
		\hat I_y + \hat N_y \\
		\hat I_z + \hat N_z
	\end{bmatrix}
	\right\|
	\label{eq:equisolid normal mapping}
\\
	\begin{bmatrix}
		\vec A_x \\
		\vec A_y
	\end{bmatrix}
	&=
	\frac{\sqrt{2}}
	{
		\left|
		\begin{bmatrix}
			\hat I_x & \hat I_y & \hat I_z+1
		\end{bmatrix}
		\right|
	}
	\begin{bmatrix}
		\hat I_x \\
		\hat I_y
	\end{bmatrix}
	&\text{, if }\hat N_z=1
	\label{eq:orthagonal equisolid mapping}
\end{align}
\end{subequations}
Inverse mapping:
\begin{subequations}
\begin{align}
	\begin{bmatrix}
		\hat A_x' \\
		\hat A_y' \\
		\hat A_z'
	\end{bmatrix}
	&=
	\begin{bmatrix}
		\vec A_x\sqrt{\nicefrac{1}{2}} \\
		\vec A_y\sqrt{\nicefrac{1}{2}} \\
		\sqrt{1-
			\frac{\vec A_x^2}{2}-
			\frac{\vec A_y^2}{2}
		}
	\end{bmatrix}
\\
	\begin{bmatrix}
		\hat I_x \\
		\hat I_y \\
		\hat I_z
	\end{bmatrix}
	&=
	2\left(
		\hat A'\cdot\hat N
		\begin{bmatrix}
			\hat A_x' \\
			\hat A_y' \\
			\hat A_z'
		\end{bmatrix}
		-
		\begin{bmatrix}
			\hat N_x \\
			\hat N_y \\
			\hat N_z
		\end{bmatrix}
	\right)
	+
	\begin{bmatrix}
		\hat N_x \\
		\hat N_y \\
		\hat N_z
	\end{bmatrix}
\\
	&=
	\begin{bmatrix}
			\vec A_x\sqrt{2-\vec A_x^2-\vec A_y^2} \\
			\vec A_y\sqrt{2-\vec A_x^2-\vec A_y^2} \\
			1-\vec A_x^2-\vec A_y^2
	\end{bmatrix}
	&\text{, if }\hat N_z=1
	\label{eq:inverse orthagonal equisolid mapping}
\end{align}
\end{subequations}
where $\vec A\in[-1,1]^2$ is the azimuthal equisolid projection coordinate. $\theta$ is the incidence angle. $\hat N\in\real^3$ is the surface normal vector. As the incidence $\hat I\in\real^3$ is mapped to or from orthogonal texture space, where $\hat N_z=1$, the transformation can take form of equation \vref{eq:orthagonal equisolid mapping} and \vref{eq:inverse orthagonal equisolid mapping}.

\begin{figure}[h]
	\begin{subfigure}[t]{\columnwidth}
		\centering
		\includegraphics{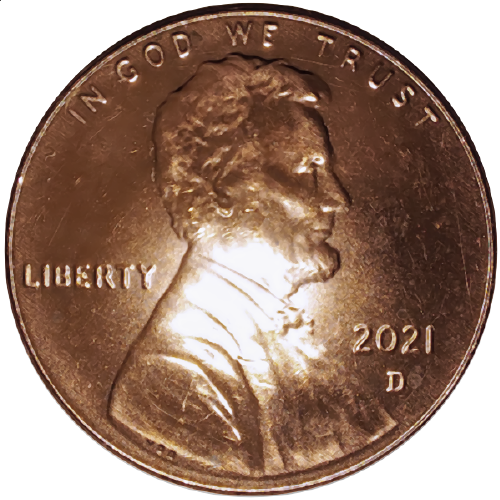}
		\caption{Picture of one cent American coin.}
	\end{subfigure}
	\hfill
	\begin{subfigure}[t]{\columnwidth}
		\centering
		\includegraphics{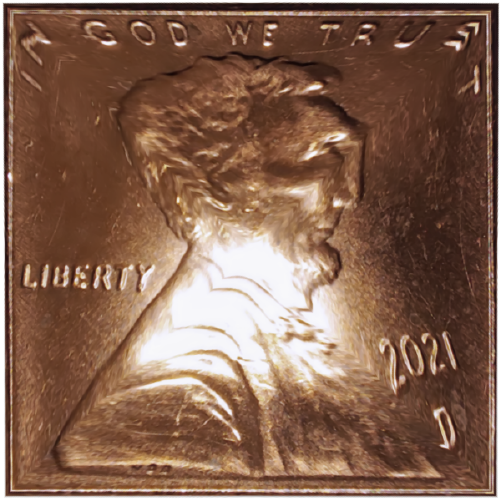}
		\caption{One cent coin mapped to a rectangle, using equation \vref{eq:incidence square mapping}.}
	\end{subfigure}
	\caption{A visual example of disc to square mapping using the formulation found in equation \vref{eq:incidence square mapping}.}
	\label{fig:square mapping}
\end{figure}

Following equation transforms azimuthal projection vector, into square coordinates, for the radiance bucket sampling.\footnote{See figure \vref{fig:square mapping} for visual reference.}
\begin{equation}
	\label{eq:incidence square mapping}
	\begin{bmatrix}
		\vec B_x \\
		\vec B_y
	\end{bmatrix}
	=
	\frac{\big|\begin{bmatrix}\vec A_x & \vec A_y\end{bmatrix}\big|}
		{\max\big(|\vec A_x|, |\vec A_y|\big)}
	\begin{bmatrix}
		\vec A_x \\
		\vec A_y
	\end{bmatrix}
	\quad\text{if }\vec A_x \text{ and } \vec A_y \neq 0 \\
\end{equation}
where $\vec B\in[-1,1]^2$ is the bucket's centered texture coordinate and $\vec A\in[-1,1]^2$ is the azimuthal projection vector.
\paragraph{Note} It is important to prevent pixel blending between edges of neighboring buckets. This can be done by clamping bucket coordinates to
$\vec B\in[B_\textit{res}^{-1}-1,1-B_\textit{res}^{-1}]^2$ range.
\medskip

Inverse transformation of bucked, centered coordinates $\vec B\in\real^2$ to azimuthal projection coordinates $\hat A\in\real^2$ can be achieved with same, but inverted method.
\begin{subequations}
\begin{align}
	\begin{bmatrix}
		\vec A_x \\
		\vec A_y
	\end{bmatrix}
	&=
	\frac{\max\big(|\vec B_x|, |\vec B_y|\big)}
		{\sqrt{\vec B_x^2+\vec B_y^2}}
	\begin{bmatrix}
		\vec B_x \\
		\vec B_y
	\end{bmatrix}
	\label{eq:outgoing incidence}
\\
	\begin{bmatrix}
		\hat R_x \\
		\hat R_y \\
		\hat R_z
	\end{bmatrix}
	&=
	\begin{bmatrix}
			-\vec A_x\sqrt{2-\vec A_x^2-\vec A_y^2} \\
			-\vec A_y\sqrt{2-\vec A_x^2-\vec A_y^2} \\
			1-\vec A_x^2-\vec A_y^2
	\end{bmatrix}
	\label{eq:reflection vector}
\end{align}
\end{subequations}
where $\hat R\in\real^3$ denotes equisolid reflection vector. This vector is used to sample ray-traced data onto radiance field texture. It is a version of the vector mirrored along the normal, found in equation \vref{eq:inverse orthagonal equisolid mapping}.

\section{Results}

\emph{TBA}

\section{Conclusion}

I have theorized about possible implementation of radiance field texturing using modern hardware shading capabilities, and presented mathematical solution for executing such concept.

\paragraph{Note} \emph{More conclusion are to be added, after the update to the paper.}

% \section{Discussion}

\section{Possible applications}

Radiance field texture sampling can replace shading pipeline or supplement it with enhanced effects. Some such effects include:

\paragraph{Parallax interior mapping}
This effect is used to mimic interior of a room, as seen through a window, or it can simulate a portal to another place.

\paragraph{Proxy meshes with parallax mapping}
Radiance texture with alpha mask can simulate more complex or furry objects bound inside a proxy mesh. Similarly to neural radiance fields texturing primitives\cite{Baatz2022NerfTex}.

\paragraph{Reflections}
Many light bounces can be combined into a single pixel of the radiance texture map. Dynamic objects can then sample such radiance field to obtain environment reflections. Also semi real-time ray-tracing can accumulate dynamically generated reflections into such texture map, to update and enhance environment one.

\paragraph{Shadowing}
1-bit radiance field texture map can represent shadowing of static objects. Here, incidence vector is replaced with light direction vector for shadow occlusion sampling. It can work with both parallel light sources and point lights. With more than one sample per bucket, area shadows are possible to produce.

\paragraph{Subsurface scattering}
This computationally demanding effect can be encoded in a radiance texture map, which then replaces incidence vector, with the light direction vector in relation to the view position for sampling.

%%%%%%%%%%%%%%%%%%
%% Bibliography %%
%%%%%%%%%%%%%%%%%%

\bibliographystyle{ACM-Reference-Format}
\bibliography{bibliography} % BibTeX format

%%%%%%%%%%%%%%
%% Appendix %%
%%%%%%%%%%%%%%

% \appendix
% \section{???}

% In this appendix,

\end{document}